\documentclass[twocolumn,preprintnumbers,amsmath,amssymb]{revtex4}

\usepackage{graphicx}
\usepackage{dcolumn}
\usepackage{bm}

\begin{document}

\title{Marathon pacing and elevation change}

\author{J. B. Elliott}
\affiliation{http://www.linkedin.com/pub/james-elliott/36/123/816}


\date{\today}
\begin{abstract}
An analysis of marathon pacing and elevation change is presented.\ \ It is based on an empirical observation of how the pace of elite and non-elite marathon runners change over the course of the marathon and a simple approximation of the energy cost of ascent and decent.\ \ It was observed that the pace of the runners slowed in a regular manner that could be broken up into four regions.\ \ That observation can be used to project target paces for a desired marathon finish time.\ \ However, that estimate fails to take in to account the energetic costs of elevation changes (hills) along the marathon course.\ \ Several approximations are made to give a coarse estimate of target paces for marathon run on courses with significant elevation changes, i.e. a hilly course.\ \ The 2012 Oakland Marathon course is used as and example of a hilly course and the times of 23 finishers are examined.
\end{abstract}

\maketitle

\section{Introduction}

This paper addresses the pace of marathons over courses with significant elevation changes (i.e. hills).\ \ The analysis starts by reporting the observed change in the pace of marathon runners on flat courses reported by Myers and Higdon \cite{higdon-book}.\ \ Here pace refers to the amount of time a runner takes to run a mile.\ \ The units, following the conventions of marathon running in the United States, are generally minutes per mile.

Then a crude analysis of the energetic costs of running hills is given.\ \ This analysis takes into account the energy cost of moving the runner both horizontally and vertically in an approximate fashion.

Next, the pace predicted by Myers and Higdon is adjusted to take into account the elevation changes of the 2012 Oakland Marathon course.\ \ An example for a marathon finishing time of three hour and thirty minute is given.\ \ The results are compared to the pace estimates for the same course given by ``smart pace bands'' from ``races2remember''\cite{races2remember}.

Finally, the paces of 23 finishers of the 2012 Oakland Marathon are examined.\ \ The average and individual splits of those 23 runners are compared to the splits predicted by the Myers hill corrected prediction based on the average finish time.\ \ Here ``split'' refers the time taken by a finisher to run a mile in the course of the marathon.

The times are generally reported in the conventional form for marathon running of hours:minutes:seconds (h:m:s) and paces and splits are reported in minutes:seconds (m:s) per mile.

\section{Myers pace regions}

In the book ``Marathon -- The Ultimate Training Guide" by Hal Higdon \cite{higdon-book}, an analysis by an engineer and marathon runner named Goerge Myers examines the pace of elite marathon runners over the course of a marathon.\ \ Myers observed that there were some standard variations in the average pace of an elite marathon finisher that could roughly be broken up into the four regions shown in Table~\ref{table1}.\ \  

\begin{table}[htdp]
\caption{Myers pace regions and factors}
\begin{tabular}{|c|c|c|c|}
\hline
Region ($n$)	& Miles	& Miles		& Factor ($f_n$)	\\
\hline
1			&$0$	&	$12$		& $0.037560$		\\
\hline
2			&$12$	&	$18$		& $0.038086$		\\
\hline
3			&$18$	&	$23$		& $0.038725$		\\
\hline
4			&$23$	&	$26.2$	& $0.039500$		\\
\hline
\end{tabular}
\label{table1}
\end{table}

The trend that Myers noted was that the pace, $p_n$, of an average marathon finisher in region $n$ could be calculated by taking the finishing time $t_f$ (in seconds) and multiplying it by the factor $f_n$ listed in Table~\ref{table1} so that:
	\begin{equation}
		p_n = t ~ f_n .
	\label{pace}
	\end{equation}
	
Table~\ref{table2} shows four examples of the Myers paces for various marathon finish times.

\begin{table}[htdp]
\caption{Myers expected paces}
\begin{tabular}{|c|c|c|c|c|}
\hline
Finish time	& $p_{n=1}$	& $p_{n=2}$	& $p_{n=3}$	& $p_{n=4}$	\\
(h:m)			&(m:s/mile)	&(m:s/mile)	&(m:s/mile)	&(m:s/mile)	\\
\hline
3:00			& 6:46		& 6:51		& 6:58		& 7:07		\\
\hline
3:15			& 7:19		& 7:26		& 7:33		& 7:42		\\
\hline
3:30			& 7:53		& 8:00		& 8:08		& 8:18		\\
\hline
3:45			& 8:27		& 8:34		& 8:43		& 8:53		\\
\hline
\end{tabular}
\label{table2}
\end{table}

Later, Myers and Higdon expanded their analysis to look at times of non-elite marathoners and found roughly the same patterns.

\begin{figure*}[htbp]
\includegraphics[width=18.0cm]{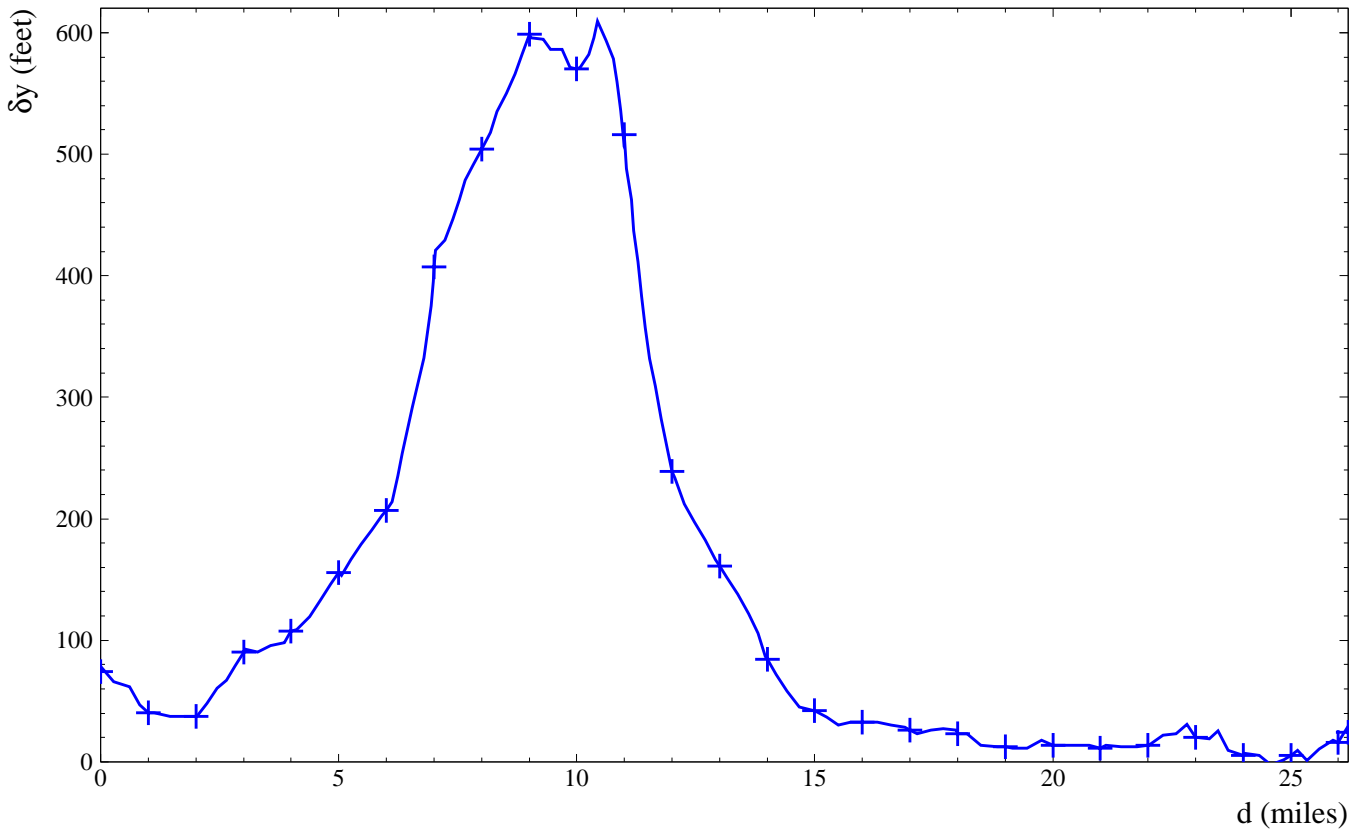}
\caption{The course elevation profile for the 2012 Oakland Marathon.\ \ The blue curve shows the course elevation profile and the blue crosses squares show the elevation at each mile.}
\label{om2012a-01}
\end{figure*}

\section{Hills}

As Higdon and Myers noted, their expected paces apply to marathons that are run on flat courses.\ \ Hills can modify a runner's pace by several minutes per mile depending on the number, duration and inclination of the hills on the course.\ \ Thus, some modification of Myers' analysis is needed for marathons with significant elevation changes.  

To account for the effects of hills, it is assumed that the energy to move a runner forward on a flat course is approximately given by
	\begin{equation}
		E_{\text{flat}} = 0.75~w~ \delta x
	\label{Eflat}
	\end{equation}
where $w$ is the weight of the runner in pounds, $\delta x$ is distance run in miles and 0.75 is a constant of proportionality that returns the energy units of nutritional calories \cite{hall-04}.\ \ 

It is then assumed that to move a runner up (or down) would require an energy given, approximately, by
	\begin{equation}
		E_{\text{hill}} = w~\delta y
	\label{Ehill}
	\end{equation}
where $\delta y$ is the elevation change.\ \ This energy can be put into calories with the appropriate constants of proportionality.\ \ For this work, $\delta y$ is given in feet. 

Finally, it is assumed that the pace of a runner, $p^*$, depends, approximately, on the pace the runner would maintain on a flat course, ($p_n$) and the energetic costs of running ($E_{\text{flat}}$ and $E_{\text{hill}}$) as follows
	\begin{eqnarray}
		p^* & \approx & p_n   \frac{E_{\text{flat}} + E_{\text{hill}}}{E_{\text{flat}}}  \nonumber \\
		& = & t ~  f_n ~  \left( 1 + \frac{\delta y}{0.75  ~ \delta x} \right) .
 	\label{pace2}
	\end{eqnarray}
Note that the weight of the runner, $w$ drops out of Eq.~(\ref{pace2}) so that it is valid for runners of any weight.\ \ On a perfectly flat course $\delta y = 0$ and $p^* = p_n$.

The energetic approximations made for both the energy expenditure on a flat course and on a hilly course are very crude, but should capture the essential energy costs.

\begin{figure}[htbp]
\includegraphics[width=8.7cm]{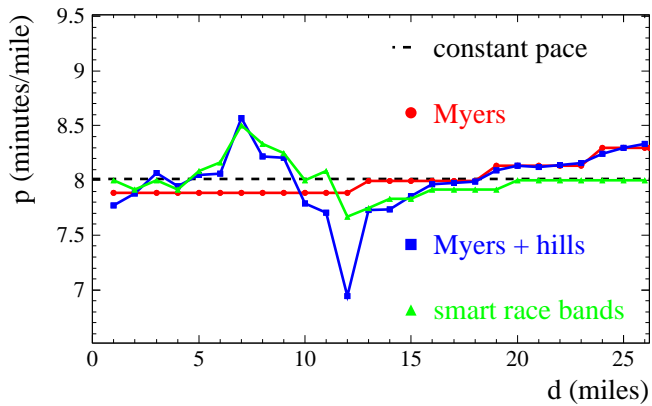}
\caption{The Myers pace for a 3:30 marathon finish time from Eq.~(\ref{pace}) is shown with red circles.\ \ The Myers pace adjusted for the hills of the 2012 Oakland Marathon via  Eq.~(\ref{pace2}) is shown with blue squares.\ \ The dotted, black line shows the constant or even pace result.\ \ The estimate of the pace for the 2012 Oakland Marathon from ``smart pace bands'' \cite{races2remember} which takes into account the hill profile, but not Myers' analysis, is shown with green triangles.}
\label{om2012-330.eps}
\end{figure}

 \section{2012 Oakland Marathon}
 
 \subsection{Expected effects of hills}

The 2012 Oakland Marathon is taken as an example of a course with significant hills.\ \ Expected paces are calculated based on the elevation changes of the course, or hill profile, shown in Fig.~\ref{om2012a-01}.

\begin{table}[htdp]
\caption{Myers expected mile splits and elapsed adjusted for hills for a 3:30 marathon finish adjusted for the 2012 Oakland Marathon elevation profile.}
\begin{tabular}{|r|c|r|}
\hline
Mile	& pace		& Elapsed time	\\
	&(m:s/mile)	&(h:m:s)	\\
\hline
1& 7:46  &  7:46	\\
2& 7:53  &  15:39	\\
3& 8:04  &  23:43	\\
4& 7:57  &  31:40	\\
5& 8:03  &  39:43	\\
6& 8:04  &  47:47	\\
7& 8:34  &  56:21	\\
8& 8:13  &  1:04:34	\\
9& 8:13  &  1:12:46	\\
10& 7:47  &  1:20:34	\\
11& 7:42  &  1:28:16	\\
12& 6:57  &  1:35:13	\\
13& 7:44  &  1:42:56	\\
14& 7:44  &  1:50:40	\\
15& 7:51  &  1:58:32	\\
16& 7:58  &  2:06:30	\\
17& 7:58  &  2:14:28	\\
18& 7:59  &  2:22:27	\\
19& 8:06  &  2:30:33	\\
20& 8:08  &  2:38:41	\\
21& 8:07  &  2:46:49	\\
22& 8:08  &  2:54:57	\\
23& 8:09  &  3:03:06	\\
24& 8:15  &  3:11:21	\\
25& 8:18  &  3:19:39	\\
26& 8:20  &  3:27:59	\\

\hline
\end{tabular}
\label{table3}
\end{table}

Over the first nine miles of the 2012 Oakland Marathon, the course climbs over 500 feet.\ \ Then, between mile 10 and mile 15 the course drops by over 500 feet.\ \ Finally, the last 11.2 miles of the course is relatively flat.\ \ With those changes in elevation, the pace predicted by Myers with Eq.~(\ref{pace}) is too fast for the first nine miles, too slow for the next six miles and approximately correct for the last 11.2 miles.\ \ A more accurate pace can be calculated with Eq.~(\ref{pace2}).

Figure~\ref{om2012-330.eps} shows the Myers pace (Eq.~(\ref{pace})) and the Myers hill corrected pace (Eq.~(\ref{pace2})) for a finish time of three hours and thirty minutes (3:30) at the 2012 Oakland Marathon.\ \ Table~\ref{table3} shows the individual mile times (splits) and elapsed time for each mile from the hill corrected Myers pace.

Also shown in Fig.~\ref{om2012-330.eps} is the pace estimate for the 2012 Oakland Marathon from ``smart pace bands'' ({\it spb})\cite{races2remember}.\ \ For each course,  {\it spb} reviews the elevation profile and adjustment the pace accordingly based on the terrain of the marathon.\ \ They consider the elevation changes, steepness, whether a mile is all up, all down or a mixture and whether the hill is early or late.\ \ Typically, {\it spb} suggests a 5--10 second decrease in pace for a 50--75 foot climb over a mile and 10--30 seconds for 100--150 feet.\ \ Once the time lost due to climbing is determined,  {\it spb} attempts to find places on the course where that time can be made up.\ \ They begin with downhills first and flat portions and increase the pace on those accordingly. 

It appears that the {\it spb} pace shown in Fig.~\ref{om2012-330.eps} is based on a constant pace.\ \ This is clear because the even splits pace (black, dotted line in Fig.~\ref{om2012-330.eps}) agrees with the {\it spb} pace after 20 miles where the 2012 Oakland Marathon course is relatively flat.

\begin{figure}[htbp]
\includegraphics[width=8.7cm]{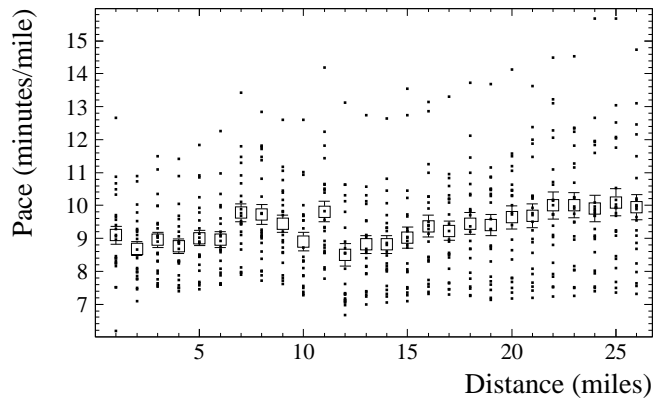}
\caption{The mile splits of 23 finishers of the 2012 Oakland Marathon.\ \ Small, solid squares show the results for individual finishers.\ \ The larger, open squares show the average pace for a given mile; the error bars show the error on the mean in the pace at a given mile.\ \ Only the first 26 miles are shown.}
\label{om2012-finishers.eps}
\end{figure}

\begin{figure}[htbp]
\includegraphics[width=8.7cm]{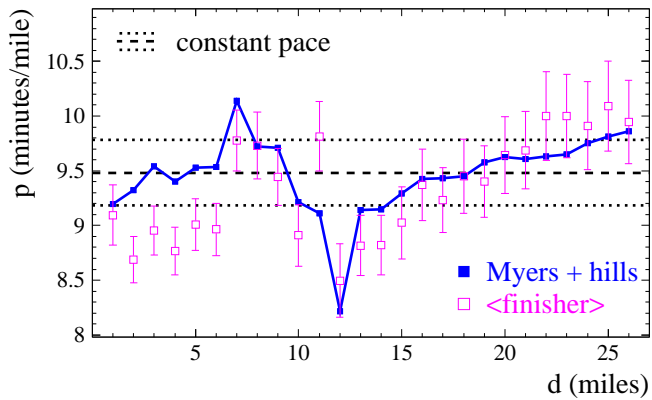}
\caption{A comparison between the average splits from 23 finishers of the 2012 Oakland Marathon and the prediction of Eq.~(\ref{pace2}).\ \ The violet squares show the average split of 23 finishers of the 2012 Oakland Marathon, the error bars show the error on the mean.\ \ The black, dashed line shows the overall average pace of the finishers, the black dotted line shows the root mean square variation.\ \ The blue squares show the predicted split based on the average time of 23 finishers of the 2012 Oakland Marathon, Eq.~(\ref{pace2}) and the 2012 Oakland Marathon elevation profile.}
\label{om2012-comp1.eps}
\end{figure}

\begin{figure*}[htbp]
\includegraphics[width=18.0cm]{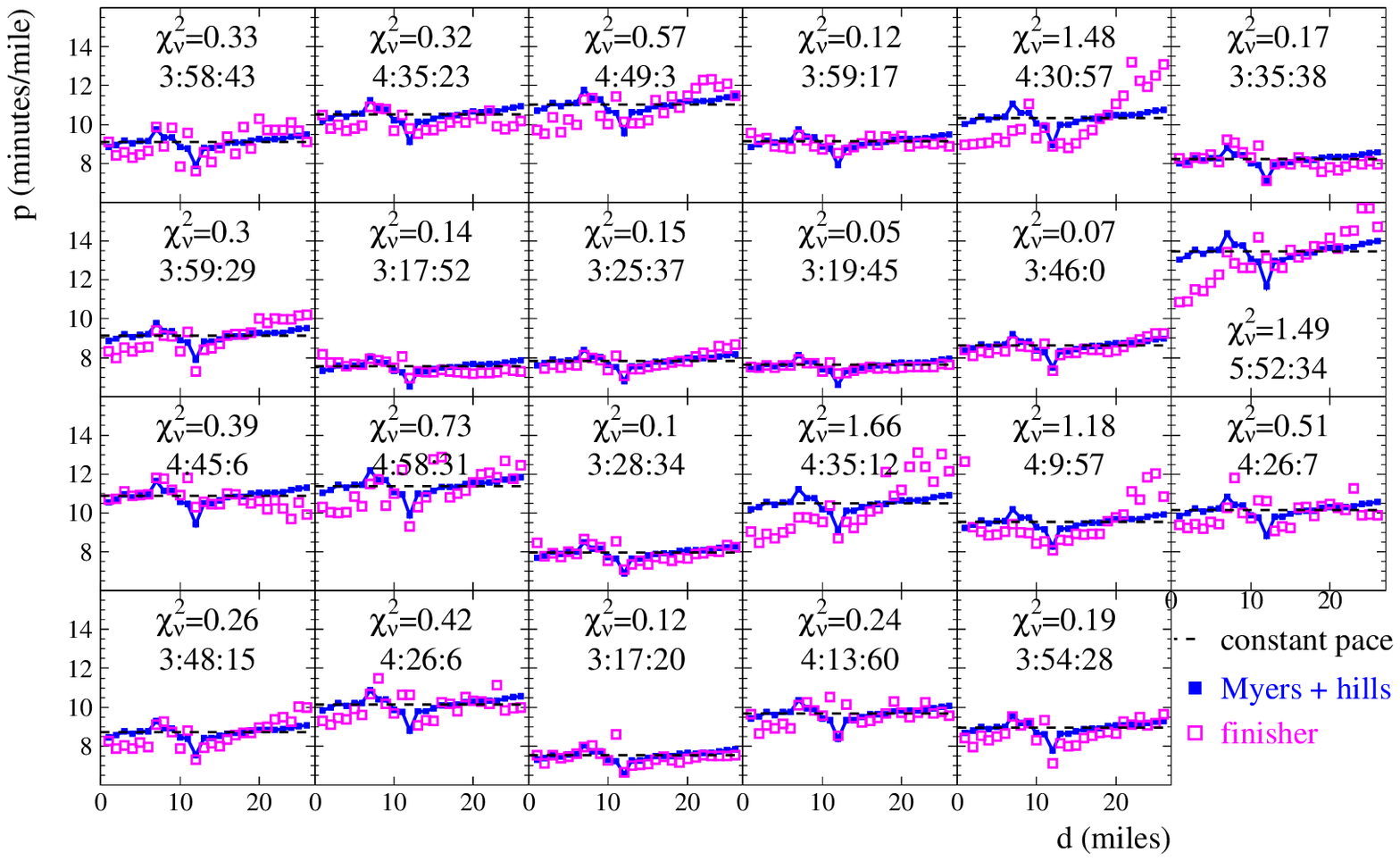}
\caption{The pace of individual finishers of the 2012 Oakland Marathon (empty, violet squares) compared the pace based on their finishing times and the Myers elevation adjusted prediction of Eq.~(\ref{pace2}) (full, blue squares).\ \ The average pace is shown with a dashed, black line.\ \ Listed on each figure is the $\chi^2_\nu$ values based on the agreement between the prediction of Eq.~(\ref{pace2}) and the actual results and the finishing times.}
\label{om2012-comp1a1.eps}
\end{figure*}

\subsection{Comparison to finshers' times}

The above analysis can be compared to the times of finishers of the 2012 Oakland Marathon to gauge the efficacy of the empirical observations of Myers and Higdon and the elevation change adjustments introduced above.

The splits from 23 finishers of the 2012 Oakland Marathon are shown in Fig.~\ref{om2012-finishers.eps}.\ \ The splits were recorded with Garmin GPS watches and posted on the Garmin Connect website \cite{garmin}.\ \ The average finishing time of those 23 finishers was 4:08:26, with a root mean square variation of 37:30 and an error on the mean of 7:49.\ \ This corresponds to an average pace of 9:29 per mile, with a root mean square variation of 1:26 and an error on the mean of 18 seconds.

The average finishing time was inserted into Eq.~(\ref{pace2}) and the elevation profile of the 2012 Oakland Marathon shown in Fig.~\ref{om2012a-01} was used to predict the pace of an average finisher.\ \ The results are shown in Fig.~\ref{om2012-comp1.eps}.

\begin{figure}[htbp]
\includegraphics[width=8.7cm]{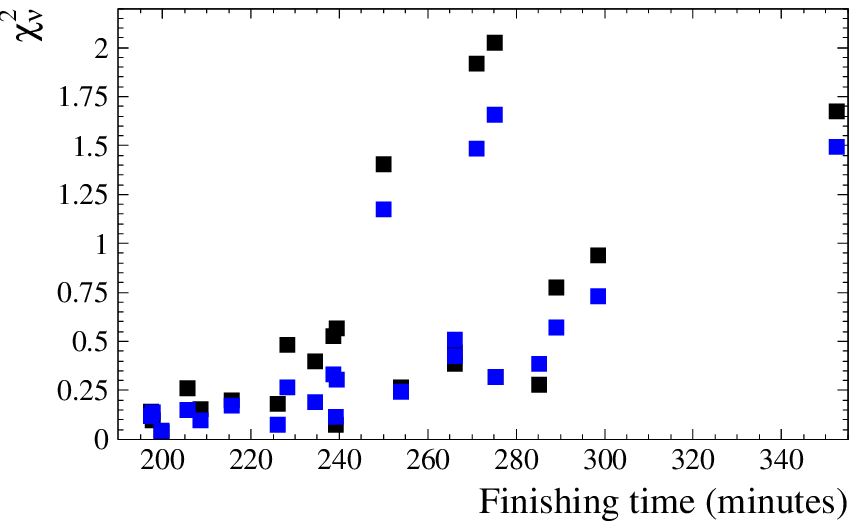}
\caption{The $\chi^2_\nu$ values based on the agreement between the prediction of Eq.~(\ref{pace2}) and the actual results for each finisher as a function of the finishing time (blue squares) and the $\chi^2_\nu$ values based on the agreement between the average, constant pace and the actual results for each finisher as a function of the finishing time (black squares).}
\label{om2012-comp1a2.eps}
\end{figure}

\begin{figure*}[htbp]
\includegraphics[width=18.0cm]{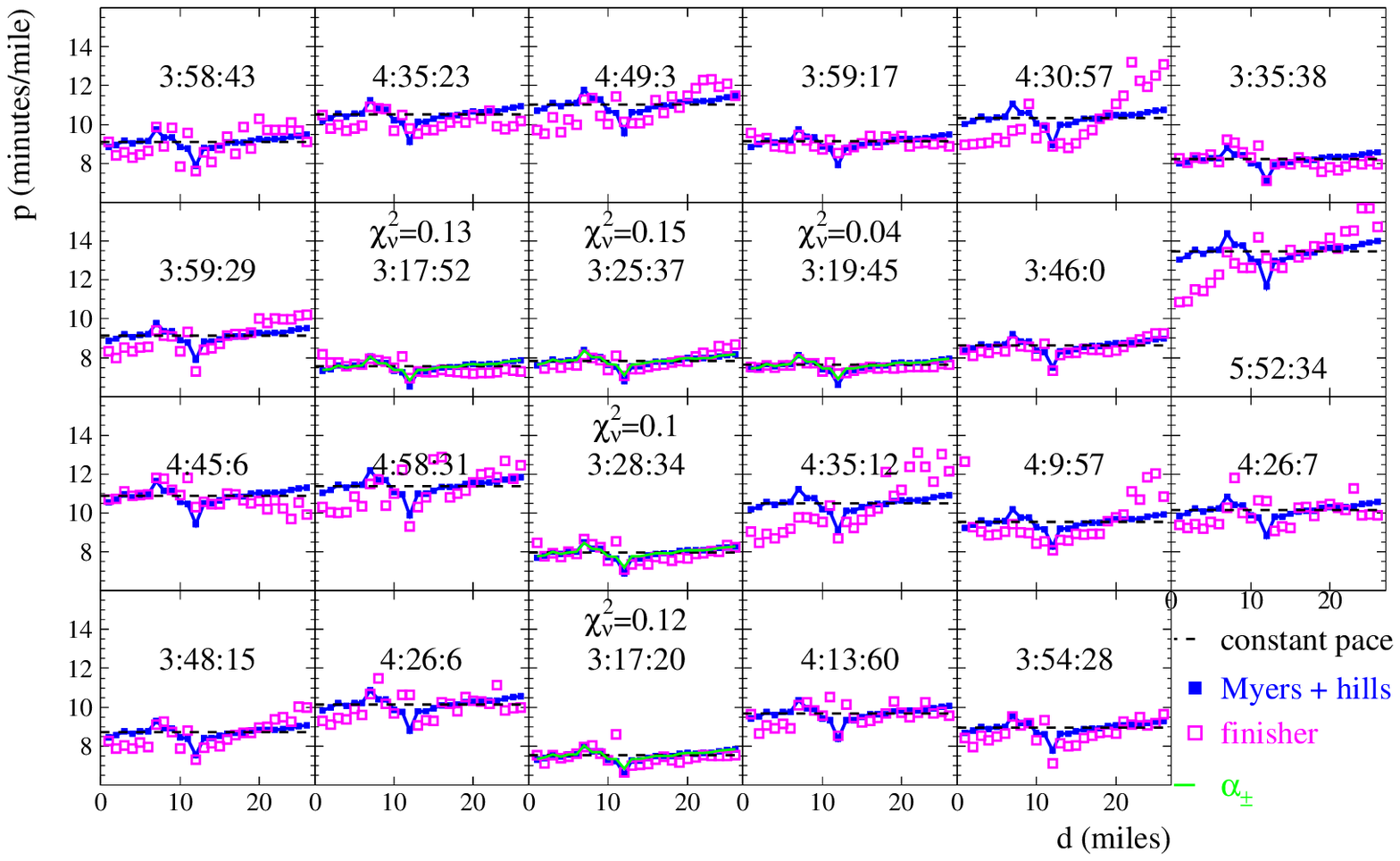}
\caption{The pace of individual finishers of the 2012 Oakland Marathon (empty, violet squares) compared the pace based on their finishing times and the Myers elevation adjusted prediction of Eq.~(\ref{pace2}) (full, blue squares).\ \ The average pace is shown with a dashed, black line.\ \ Listed on each figure is the $\chi^2_\nu$ values based on the agreement between the fit of Eq.~(\ref{pace3}) and the actual results and the finishing times.\ \ The solid, green line shows the fit.}
\label{om2012-comp1a1s.eps}
\end{figure*}

Figure~\ref{om2012-comp1.eps} shows that the paces predicted with Eq.~(\ref{pace2}) for miles two through six and mile 11 do not agree with the finishers' average pace.\ \ The source of the discrepancy observed for miles two through six is discussed below\ \ The discrepancy observed at mile 11 is also addressed below.\ \ However, the paces predicted for all other miles agree, to within error bars, with the observed average paces of the finishers.\ \ Comparing the Myers hill adjusted prediction compared to the 23 2012 Oakland Marathon finishers' results yields $\chi^2_\nu = 2.0$ with no free parameters.

Comparing the average finisher splits over the entire marathon with the constant pace mile split yields $\chi^2_\nu = 3.1$.\ \ This shows that the Myers elevation adjusted pace more accurately reflects the pace at each mile than the constant pace over the entire marathon.\ \ That is, given a finishing time and the elevation profile of a marathon course, Eq.~(\ref{pace2}) can more accurately predict the time taken to run individual miles along the course than just determining the average pace based on the finishing time divided by 26.2 miles.

The finishing time of the individual runner's can also be compared to Eq.~(\ref{pace2}) directly.\ \ This is shown in Fig.~\ref{om2012-comp1a1.eps}.\ \ A visual inspection of Fig.~\ref{om2012-comp1a1.eps} shows that the discrepancy for miles two through six observed in Fig.~\ref{om2012-comp1.eps} is more pronounced for the slower finishers who ran the initial portion of the marathon faster than the predictions of  Eq.~(\ref{pace2}) and finished running slower than those predictions.\ \ In the vernacular of marathon runners, the finishers with more pronounced disagreements with the predicted splits of  Eq.~(\ref{pace2}) {\it took it out to hard and then crashed}.

Figure~\ref{om2012-comp1a2.eps} shows the $\chi^2_\nu$ values of the splits predicted by  Eq.~(\ref{pace2}) and the actual splits.\ \ That plot shows that the splits of runners with faster finishing times are better described by  Eq.~(\ref{pace2}), while the splits of the slower finishers are less accurately described.\ \ Considering all the splits of all the finishers together gives $\chi^2_\nu = 0.48$.\ \ This is lower than the $\chi^2_\nu$ for the comparison in Fig.~\ref{om2012-comp1.eps} because in Fig.~\ref{om2012-comp1a1.eps},  Eq.~(\ref{pace2}) was applied to each finisher individually.

Also shown in Fig.~\ref{om2012-comp1a2.eps} are the $\chi^2_\nu$ values of the splits based on a constant pace given by the finisher's time divided by 26.2 miles and the actual splits.\ \ This shows that, on average,  Eq.~(\ref{pace2}) more accurately predicts the individual mile times than does the simple constant pace average with the average difference in $\chi^2_\nu$ being 0.11 in favor of the predictions of  Eq.~(\ref{pace2}), an 18\% improvement in $\chi^2_\nu$.

\subsection{Improvements}

One problem with the approximations inherent in the analysis presented above is that the human body is not as simple as a ball rolling down hill.\ \ As a human runs down hill, the runner's kinetic energy is not increased by an amount equal to the loss in potential energy due to the elevation change.\ \ This is the result of a myriad of inelastic collisions and compression in the human body.

To account for this in a very approximate fashion,  Eq.~(\ref{pace2}) can be rewritten as:
	\begin{equation}
		p^* \approx  t ~  f_n ~  \left( 1 + \frac{\alpha_{\pm}  ~ \delta y}{0.75 ~  \delta x} \right) .
 	\label{pace3}
	\end{equation}
where $\alpha_{\pm}$ is some factor with one value if $\delta y \ge 0$ (flat or uphill running) and another for $\delta y < 0$ (downhill running).

Figure~\ref{om2012-comp1a1s.eps} shows the results of fitting the splits to finishers who finished under 3:30:00 in order to avoid the problem of {\it taking it out too hard} mentioned above.\ \ The fit gives $\alpha_+ = 0.9\pm0.2$ and $\alpha_- = 0.6\pm0.1$ with $\chi^2_\nu = 0.106$.\ \ When the values of $\alpha_{\pm} = 1$ was held fixed, $\chi^2_\nu = 0.111$ for this subset of the data, so the fit parameters yield a 4.5\% lower value of $\chi^2_\nu$.

\begin{table}[htdp]
\caption{Comparisons of predicted pace for mile 11 for runners finishing the 2012 Oakland Marathon under 3:30.}
\begin{tabular}{|c|c|c|c|}
\hline
Finisher	& $\Delta$pace			& $\Delta$pace			& Improvement	\\
Number	& Eq.~(\ref{pace2})	& Eq.~(\ref{pace3})	&			\\
		&(m:s/mile)			&(m:s/mile)			& (m:s)/mile	\\
\hline
8	& 0:47	& 0:41	& 0:06	\\
9	& 0:34	& 0:27	& 0:07	\\
10	& 0:26	& 0:20	& 0:06	\\
15	& 0:54	& 0:47	& 0:07	\\
21	& 1:23	& 1:16	& 0:07	\\
\hline
\end{tabular}
\label{table4}
\end{table}

The values of $\alpha_{\pm}$ suggest that the energy loss due to compression effects of the human body is small (and consistent with zero) during flat or uphill running.\ \ However, while running downhill the energy loss is approximately $40\pm10$\%.

This phenomenon should explain the discrepancy at mile 11 noted above and shown in Fig.~\ref{om2012-comp1.eps}.\ \ From mile 10 to mile 11 there is a net decrease in elevation and this net decrease is the only factor in determining the elevation adjustment to the pace as determined via  Eq.~(\ref{pace2}); the actual path of the runner up and down over the course of a mile does not matter.\ \ However, if  Eq.~(\ref{pace3}) is used, then the actual path of the runner up and down does matter since $\alpha_{\pm}$ breaks the up and down symmetry.

Unfortunately, Fig.~\ref{om2012-comp1a1s.eps} shows that the fits using  Eq.~(\ref{pace3}) fail to match the actual finshers' splits for mile 11.\ \ Table~\ref{table4} shows the deviation in the predicted pace and the actual pace, $\Delta$pace, for the two methods.\ \ Using  Eq.~(\ref{pace3}) instead of  Eq.~(\ref{pace2}) does lead to a 6.6 second decrease in the deviation, but that is only 13.5\% of the observed deviation.

The deviation in the observed and predicted pace of mile 11 is an open question.\ \ It may be possible that the elevation profile used here (and shown in Fig.~\ref{om2012a-01}) is inaccurate between mile ten and mile 11.\ \ Some evidence for this can be seen in Fig.~\ref{om2012-330.eps} where the {\it spb} estimate for the pace of mile 11 (filled green triangles and green line) is slower than the pace of mile ten.\ \ This is the opposite of the pace estimate from  Eq.~(\ref{pace2}) (blue squares and blue line) in Fig.~\ref{om2012-330.eps}.\ \ Other than tha the first mile, mile 11 and after mile 12 (where  Eq.~(\ref{pace2}) takes into account the pace slowing effects that Myers noted, which was not taken into account by the {\it spb} estimate) the trend of the pace change from one mile to the next is similar between the two estimates.

\section{Summary}

An approximate treatment of the effects of elevation changes on the pace of marathon runners was presented.\ \ It was seen that a good approximation of 23 finishers of the 2012 Oakland Marathon was made using  Eq.~(\ref{pace2}).\ \ Some crude corrections for compression effects of the human body were made leading to an improved agreement between  Eq.~(\ref{pace3}) and the faster finishers.\ \ This analysis can be used to plan a strategy for runners of marathon courses with significant elevation changes (hills).

Improvements to the analysis presented here could be achieved with a better approximation to the energy cost of flat and uphill and down hill running.\ \ More data consisting of the splits of more runners on more marathon course would also improve estimates of $\alpha_{\pm}$.

\end{document}